\documentclass[aps,prb,showpacs,twocolumn,groupedaddress,floatfix]{revtex4}
\usepackage{graphicx}
\usepackage{color}
\usepackage{amsmath}
\begin{document}

\title{
Quenching of magnetic excitations in single adsorbates at surfaces:
Mn on CuN/Cu(100)
}
\author{Frederico D. Novaes$^{1}$}
\author{Nicol\'as Lorente$^{2}$}
\author{Jean-Pierre Gauyacq$^{3,4}$}
 \affiliation{
$^{1}$ Institut de Ci\`encia de Materials de Barcelona (CSIC),
Campus de la UAB, E-08193 Bellaterra, Spain\\
$^{2}$ Centre d'Investigaci\'o en Nanoci\`encia  i
Nanotecnologia (CSIC-ICN), Campus de la UAB, E-08193 Bellaterra, Spain \\
$^3$ CNRS, Institut des Sciences Mol\'eculaires d'Orsay, ISMO, Unit\'e de Recherches CNRS-Universit\'e Paris-Sud,
 B\^atiment 351, Universit\'e Paris-Sud, 91405 Orsay CEDEX,
France\\
$^4$ Universit\'e Paris-Sud, Institut des Sciences Mol\'eculaires d'Orsay, ISMO, Unit\'e de Recherches CNRS-Universit\'e Paris-Sud, B\^atiment 351, Universit\'e Paris-Sud, 91405 Orsay CEDEX,
France\\
}
\date{\today}

\begin{abstract}
The lifetimes of spin excitations of Mn adsorbates on CuN/Cu(100) are computed from first-principles. The theory is based on a strong-coupling T-matrix approach that evaluates the decay of a spin excitation due to electron-hole pair creation. Using a previously developed theory [Phys. Rev. Lett. {\bf 103}, 176601 (2009) and Phys. Rev. B {\bf 81}, 165423 (2010)], we compute the excitation rates by a tunneling current for all the Mn spin states. A rate equation approach permits us to simulate the experimental results by Loth and co-workers [Nat. Phys. {\bf 6}, 340 (2010)] for large tunnelling currents, taking into account the finite population of excited states. Our simulations give us insight into the spin dynamics, in particular in the way polarized electrons can reveal the existence of an excited state population. In addition, it reveals that the excitation process occurs in a way very different from the deexcitation one. Indeed, while excitation by tunnelling electrons proceeds via the s and p electrons of the adsorbate, deexcitation mainly involves the d electrons.
\end{abstract}


\maketitle

\section{Introduction}

Recently, a series of experimental
studies~\cite{Hirjibehedin06,Hirjibehedin07,Tsukahara,XiChen,Iacovita,Fu}
with low-temperature STM (Scanning Tunnelling Microscope) revealed that
isolated adsorbates on a surface could exhibit a magnetic structure,
i.e. that a local spin could be attributed to the adsorbate. Interaction
of this local spin with its environment results in several magnetic
energy levels that correspond to different orientations of the
local spin relative to the substrate.  A magnetic field, B, was also
applied to the system: increasing the B field decouples the local spin
from its environment and the system switches to a Zeeman structure,
thus helping to characterize the adsorbate magnetic anisotropy. The
energy of the various magnetic levels were obtained via a low-T IETS
(low-Temperature Inelastic Electron Tunnelling Spectroscopy) experiment,
in which the tip-adsorbate junction conductivity as a function of the STM
bias exhibits steps at the magnetic excitation thresholds. The magnetic
excitation energies are small, typically in the few meV range. Besides
the spectroscopy properties, the IETS experiments also revealed that the
conductance steps at the magnetic inelastic thresholds are very high,
i.e. that the efficiency of the tunneling electrons in inducing magnetic
transitions in the adsorbate is extremely large. In Metal-Phthalocyanine
adsorbates for example~\cite{Tsukahara,XiChen}, the tunneling current
is dominated by its inelastic component when the bias is above the
magnetic excitation thresholds. This efficiency is at variance with
the efficiency of tunneling electrons in exciting vibrational modes
in a molecular adsorbate which was observed and shown to only reach
the few \% range~\cite{Ho,Komeda,Lorente00}. Several theoretical
accounts of the magnetic excitation process have been reported based on
perturbation theory~\cite{Hirjibehedin07,Fransson,Fernandez,Persson},
or on a strong coupling approach~\cite{Lorente09,Gauyacq10}. In
particular, the strong coupling approach quantitatively accounted for
the very large efficiency of tunneling electrons in inducing magnetic
transitions~\cite{Lorente09,Gauyacq10}.

The existence on a surface of nano-magnets, the orientation of which
could be changed at will by tunneling electrons, opens fascinating
perspectives for the miniaturization of electronics. However, to lead
to easily manageable devices, the excitation of local spins must have,
among other properties, a sufficiently long lifetime. It is thus of
paramount importance to know the decay rate of the excited levels of
the local spin and in particular to decipher the various parameters
and effects that govern its magnitude. Experimentally, the local
spins were observed in systems in which a coating on the surface was
separating the magnetic adsorbate carrying the local spin from the metal
substrate. Experiments on adsorbates directly deposited on a metallic
substrate did not lead to sharp IETS structures~\cite{Balashov09} as the
others and this was attributed to a too short lifetime of the magnetic
excitation on metals, stressing the importance of the decoupling
layer between local spin and substrate in stabilizing the magnetic
excitation. Deexcitation of a local spin implies an energy transfer
from the local spin to the substrate degrees of freedom, i.e. to the
substrate electrons or to phonons. Phonons are not directly coupled to
spin variables, but only via spin-orbit couplings (see e.g. a discussion
in [\onlinecite{Fabian99}]). In contrast, the adsorbate spin variables
can be directly coupled to substrate electrons and electrons colliding on
a magnetic adsorbate can easily induce magnetic transitions. Actually,
this is exactly what happens in the magnetic excitation induced by
tunneling electrons in the IETS experiments described above; in the
de-excitation process the tunneling electrons are simply replaced by
substrate electrons. The decay of excited magnetic states in individual
adsorbates thus proceeds via electron-hole pair creation. Substrate
electrons colliding on the adsorbate can be thought to be as efficient
in inducing magnetic transitions as tunneling electrons injected from an
STM tip. In this qualitative view, one can expect the magnetic excitation
decay rate to be the product of the collision rate of substrate electrons
on the adsorbate by a very high efficiency factor.

Recently, the decay rate of excited magnetic Mn atoms adsorbed on
CuN/Cu(100) has been measured by Loth et al~\cite{Loth10}
via the analysis of the dependence of the adsorbate conductivity
on the tunneling current. The decay of the magnetic excitations
was interpreted in the above scheme as a decay induced by collision
with substrate electrons. The lifetimes of magnetic excitations were
typically found to be of the order of a fraction of ns. In the present
paper, we report on a theoretical ab initio study of the lifetime of
magnetic excitations in the Mn/CuN/Cu(100) system using both a DFT-based
(Density Functional Theory) description of the system and the strong
coupling formalism~\cite{Lorente09,Gauyacq10} developed to treat magnetic
transitions induced by tunneling electrons; the corresponding results
are compared with Loth et al data~\cite{Loth10}.

\section{Method}
\label{Method}
\subsection{Description of the magnetic deexcitation}
\label{Description}

The present treatment of the decay of magnetic excitations closely
parallels our earlier treatment of magnetic excitations in IETS (see
details in Ref.  [\onlinecite{Gauyacq10}]). We assume that the magnetic
levels of the adsorbate can be described by the following magnetic
anisotropy Hamiltonian~\cite{Yosida96},

\begin{equation}
        H = g \mu_{B} \vec{B} \cdot \vec{S}  + D S_{z}^{2} + E (S_{x}^{2}-S_{y}^{2}),
\label{hamiltonien}
\end{equation}
where $\vec{S}$  is the local spin of the
adsorbate, $g$ the Land\'e factor and $\mu_B$ the Bohr magneton.
$\vec{B}$ is an applied magnetic field. $D$ and $E$ are two energy
constants describing the interaction of $\vec{S}$ with the substrate,
i.e. the magnetic anisotropy of the system. Diagonalisation
of Hamiltonian~(\ref{hamiltonien}) yields the various states,
$|\Phi_i\rangle$, of the local spin of Mn (we have used S=2.5,
g=1.98, D=-41 $\mu$eV and E=7 $\mu$eV obtained in the experimental
work~\cite{Loth10} by adjustment to the magnetic excitation energy
spectrum). For Mn on CuN, the local spin is 2.5 and there are thus 6
magnetic levels.

\begin{figure}
\includegraphics[angle=0,width=0.4\textwidth]{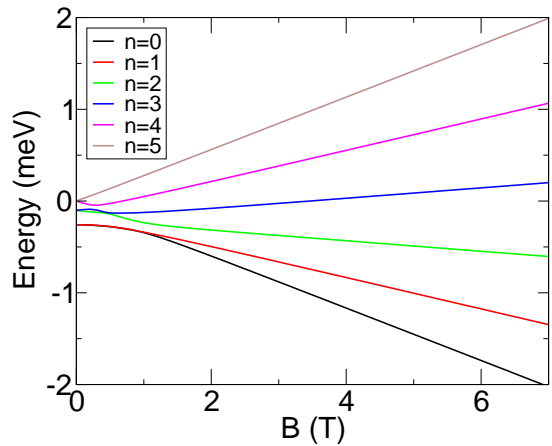}
\caption{
Energies of the six magnetic states in the Mn/CuN/Cu(100) system, eigenvalues of the magnetic anisotropy Hamiltonian ~(\ref{hamiltonien}). They are presented as a function of the applied magnetic field, B, for a field along the x-direction.
}
\label{fig_Mag_En_Lev}
\end{figure}

Figure \ref{fig_Mag_En_Lev} presents the energies
 $E_i$ of the anisotropy
states, eigenstates of Hamiltonian (\ref{hamiltonien}) as a function of
B, the applied magnetic field. Below, the ground state is noted ‘0’
and the excited states with i = 1-5. The easy axis of the system, the
z-axis, is normal to the surface in this system. In the present study,
coherently with the experimental study of Loth et al~\cite{Loth10}, the B
field has been put parallel to the surface, along the x-axis. As B is
increased, the magnetic structure of the system changes from a magnetic
anisotropy induced by the substrate with three doubly degenerate states at
B=0 to the six states of a quasi-Zeeman structure at large B, where the
$|\Phi_i\rangle$ states are eigenstates of $\vec{S}^2$ and $S_{x}$. The
ground state at large finite B corresponds approximately to the $M_x=-2$
state. The energy diagram in Fig. \ref{fig_Mag_En_Lev} thus corresponds to
the decoupling of the magnetic anisotropy by the B field. The structure
appears a little complex at low B with several avoided crossings since
the B field is not along the principal magnetic axis of the system.

The aim of the present work is to compute the decay of the excited
states by collision with substrate electrons, i.e. by electron-hole pair
creation. The decay rate, $\Gamma_{Tot,i}$, of an excited state, $|\Phi_i\rangle$, eigenstate
of the Hamiltonian ~(\ref{hamiltonien}) with energy $E_i$, is the inverse
of its lifetime $\tau_i$ and it can then be written using matrix elements
of the T transition matrix (we assume the energy variation of the T matrix
to be small on the energy scale of the $i \rightarrow f$ transition and
we assume a vanishing temperature of the substrate)~\cite{Lorente08},

\begin{eqnarray}
 \frac{1}{\tau_i} &=& \Gamma_{Tot,i} = \sum_f \Gamma_{i,f} = \sum_f \frac{2\pi\delta\Omega_f}{\hbar} \nonumber \\ 
&\times& \sum_{\substack{k_i,k_f \\ m_i,m_f}} \left|\langle 
k_f,m_f,\Phi_f |\hat{T}| k_i,m_i,\Phi_i \rangle\right|^2  \nonumber \\
&\times& \delta(\varepsilon_i-\varepsilon_f) \delta(\varepsilon_i-E_F),
\end{eqnarray}
where $|\Phi_i\rangle$ are the final states of the decay, associated to
an energy transfer of $\delta\Omega_f=E_i-E_f$, and the total energy is
$E_T=E_i+\varepsilon_i=E_f+\varepsilon_f$. The initial and final states of
the substrate electrons are noted by their wave numbers, $k_i$ and $k_f$,
and by their initial and final spin projections on the quantization axis,
$m_i$ and $m_f$. The substrate is assumed to be non-magnetic. Each term,
$\Gamma_{i,f}$, in the sum over $f$ is the partial decay rate of the
initial state to a peculiar final state.

The electron-adsorbate collisions being very fast, we can treat them
without the magnetic anisotropy (Hamiltonian ~(\ref{hamiltonien})) taken
into account and later, include it in the sudden approximation. The
electron-adsorbate collision is treated in a DFT-based approach (see
section \ref{DFT-Model} below) whereas the spin transitions are treated
in the sudden approximation in a way similar to our earlier study
of magnetic excitation~\cite{Lorente09,Gauyacq10}. We thus define a
collision amplitude for the substrate electrons independently of the
magnetic anisotropy, it is a function of both the initial and final
electron momenta, $k_i$ and $k_f$, and of the spin coupling between
electron and adsorbate. The electron-adsorbate spin coupling scheme is
defined via, $\vec{S}_T$, the total spin of the electron-adsorbate system:
$\vec{S}_T=\vec{S}+\vec{s}$, where $\vec{S}$ is the adsorbate spin and
$\vec{s}$ is the tunneling electron spin. The projection of $\vec{S}_T$ 
on the quantization axis is noted $M_T$. If the adsorbate spin is $S$,
there are two collision channels corresponding to $S_T=S+\frac{1}{2}$
and $S_T=S-\frac{1}{2}$. The decay rate is then expressed as a function
of the two channel amplitudes, $T^{S_T}$:

\begin{widetext}
\begin{eqnarray}
 \frac{1}{\tau_i}
&=& \sum_f \frac{2\pi\delta\Omega_f}{\hbar} \sum_{\substack{k_i,k_f \\ m_i,m_f}}  
\left|\sum_{S_T,M_T} \langle k_f,m_f,\Phi_f|S_T,M_T\rangle T^{S_T}\langle S_T,M_T|k_i,m_i,\Phi_i \rangle\right|^2 
\delta(\varepsilon_i-\varepsilon_f) \delta(\varepsilon_i-E_F),
\end{eqnarray}
that can be re-expressed as:
\begin{eqnarray}
 \frac{1}{\tau_i} 
&=& \sum_f \frac{2\pi\delta\Omega_f}{\hbar} \sum_{\substack{k_i,k_f \\ m_i,m_f}} \delta(\varepsilon_i-\varepsilon_f) \delta(\varepsilon_i-E_F)
\left|\sum_{S_T} \langle k_f|T^{S_T}|k_i\rangle \sum_{M_T}\langle m_f,\Phi_f|S_T,M_T\rangle \langle S_T,M_T|m_i,\Phi_i \rangle\right|^2.
\end{eqnarray}
\end{widetext}
One can see that the contributions from the two $S_T$
terms are interfering. This corresponds to the {\em a priori} general
case where no selection rule apply to the tunnelling process. In
that case, the above expresion can be simplified by making an extra
statistical approximation that neglects the interferences between the
two $S_T$ channels. However, in our earlier studies on excitation
processes~\cite{Lorente09,Gauyacq10} it was  found that one of the
two $S_T$ channels was dominating the tunneling process between tip
and adsorbate. Here for deexcitation in the Mn on CuN/Cu(100) system,
only one $S_T$ coupling scheme  does contribute significantly to
the collision between substrate electrons and adsorbate (see Section
\ref{DFT-Model}) and is thus included in the present treatment. One can
note though that the dominating channel for the deexcitation process
(collisions with electrons coming from the substrate) is different from
that dominating the excitation (collision with tunnelling electrons)
(see Section \ref{DFT-Model}).

If one considers a single $S_T$ channel, the decay rate can be rewritten as:

\begin{widetext}
\begin{eqnarray}
 \frac{1}{\tau_i} 
&=& \sum_f \frac{2\pi\delta\Omega_f}{\hbar} \sum_{\substack{k_i,k_f \\ m_i,m_f}} \delta(\varepsilon_i-\varepsilon_f) \delta(\varepsilon_i-E_F)
 \left|\langle k_f|T^{S_T}|k_i\rangle\right|^2 \left|\sum_{M_T}\langle m_f,\Phi_f|S_T,M_T\rangle \langle S_T,M_T|m_i,\Phi_i \rangle\right|^2 
\nonumber \\ 
&=& \sum_f \frac{2\pi\delta\Omega_f}{\hbar} \sum_{\substack{k_i,k_f \\ m_i,m_f}} \delta(\varepsilon_i-\varepsilon_f) \delta(\varepsilon_i-E_F)
 \left|\langle k_f|T^{S_T}|k_i\rangle\right|^2 P^{S_T}_{i,m_i\rightarrow f,m_f},
\label{eq_av_chann}
\end{eqnarray}
where $P^{S_T}_{i,m_i\rightarrow f,m_f}$  is a matrix element of spin
variables only, corresponding to the $S_T$ coupling scheme for the
$(i,m_i \rightarrow f, m_f)$ transition. Equation (\ref{eq_av_chann})
then becomes:

\begin{eqnarray}
 \frac{1}{\tau_i} &=& \sum_f \frac{2\pi\delta\Omega_f}{\hbar}
  \left(\sum_{k_i,k_f} \delta(\varepsilon_i-\varepsilon_f) 
\delta(\varepsilon_i-E_F) \left|\langle  k_f|T^{S_T}|k_i\rangle\right|^2  \right) \left(\sum_{m_i,m_f} P^{S_T}_{i,m_i\rightarrow f,m_f} \right) 
\nonumber \\
&=& \sum_f \frac{2\pi\delta\Omega_f}{\hbar}
 T^{S_T}(E_F) \left(\sum_{m_i,m_f} P^{S_T}_{i,m_i\rightarrow f,m_f} \right)
\label{eq_av_chann_short}
\end{eqnarray}
The total and partial decay rates are then obtained via:

\begin{eqnarray}
 \frac{1}{\tau_i} =  \sum_f \Gamma_{i,f} = \sum_f \frac{\delta\Omega_f}{h} 
  (2\pi)^2 T^{S_T}(E_F) P_{Spin}(S_T,i\rightarrow f).
\end{eqnarray}
\end{widetext}

Each partial decay rate thus appears as the  product of
a spin transition probability by
the electron flux hitting the adsorbate in the energy range able to
perform the studied transition.

 As we
will see below, the DFT-based calculation of the electron flux is not
performed in the $S_T$, $M_T$ spin coupling base, but by specifying the
scattering electron spin state (majority or minority). Thus we cannot
perform a one to one identification. 
However, scattering through the adsorbate being dominated by a
$S_T$ channel , we can identify the electron flux in the $S_T$ channel ($T^{S_T}(E_F)$) 
with its equivalent in the DFT-approach, the total electron flux hitting
 the adsorbate $T^{Total}(E_F)$. So that finally, the total decay rate is obtained as:

\begin{eqnarray}
 \frac{1}{\tau_i} &=&  \Gamma_{Tot,i} \nonumber \\ &=& \sum_f \Gamma_{i,f} 
 = T^{Total}(E_F) \sum_f \frac{\delta\Omega_f}{h} {P}_{Spin}(i\rightarrow f)
\label{eq_decay_final}
\end{eqnarray}

The decay rate is then equal to the total flux of substrate
electrons hitting the adsorbate per second in the appropriate energy
range, $T^{Total}(E_F)\delta\Omega_f$, times a spin transition
probability. Below, 
 $T^{Total}(E_F)$ is identified with the equivalent
quantity computed in the DFT approach (Section \ref{DFT-Model}). One can
stress the great similarity of this expression with that of the inelastic
conductivity obtained in Ref. [\onlinecite{Lorente09,Gauyacq10}] as
the total conductivity times a spin transition probability.

\subsection{DFT-based calculation of the substrate electron collision rate}
\label{DFT-Model}

The quantity that we want to compute from first-principles simulations is,

\begin{equation}
 T(E_F) = (2\pi)^2\sum_{k_i,k_f} \left| \langle k_i|\hat{T}|k_f \rangle \right|^2 \delta(\varepsilon_i-\varepsilon_f)\delta(\varepsilon_i-E_F)
\label{eq_T_ST}
\end{equation}
In this equation, $|k_i\rangle$ and $|k_f\rangle$ are asymptotic states
that are solutions to the Hamiltonian sufficiently \textit{far away} from
a scattering center. In our case, the scattering center is the adsorbed
atom. The goal is to calculate the amount of substrate (CuN/Cu(100)
surface) electrons that scatter at the adsorbed Mn atom. From this quantity, we obtain $T^{Total}(E_F)$ by summing the values for both spins,

\begin{equation}
T^{Total}(E_F)=T(E_F,\uparrow)+T(E_F,\downarrow)
\end{equation}

Following standard scattering theory formalism, we write the
solutions to the full Hamiltonian (surface+adsorbed atom) as,

\begin{equation}
 |\psi\rangle = |k_i\rangle + \hat{G}^r\hat{V}|k_i\rangle = |k_i\rangle + \hat{G}^{0r}\hat{V}|\psi\rangle
\label{eq_LS}
\end{equation}
where $\hat{V}$ is the perturbation that couples the adsorbed atom and
the substrate; and $\hat{G}^r$ and $\hat{G}^{0r}$ are the full system's
and unperturbed retarded Green's functions, respectively. The $\hat{T}$
operator is then written as,

\begin{equation}
 \hat{T} = \hat{V} + \hat{V}\hat{G}^r\hat{V}
\label{eq_T_op}
\end{equation}

In order to actually compute these vectors, matrices, and ultimately
$T(E_F)$, we have used the SIESTA\cite{Siesta} code, and a set of
strictly localized atomic-like orbitals as a basis set. In terms of these,
the Hamiltonian has the form,

\begin{equation}
 \mathbf{H}=\left(\begin{array}{cc} \mathbf{H}_{SS} & \mathbf{H}_{SA} 
      \\ \mathbf{H}_{AS} & \mathbf{H}_{AA} 
      \end{array}\right)
\label{eq_DFT_Mat}
\end{equation}
where the $\mathbf{H}_{\mu\mu}$ are themselves matrices whose elements
$H_{i,j}=\langle \phi_i|H^{DFT}|\phi_j\rangle$ are computed with orbitals
centered around either surface (S) atoms or around the adsorbed atom
(A). These orbitals are not orthogonal, $\langle \phi_i|\phi_j\rangle =
S_{i,j}$. In this representation, the coefficients of the $|\psi\rangle$ and $|k_i\rangle$ states form a vector of components~\cite{Deson_Transp},

\begin{equation}
 \mathbf{c}_{\psi} = \left(\begin{array}{c} \mathbf{c}_S \\ \mathbf{c}_A \end{array}\right)
\end{equation}

\begin{equation}
 \mathbf{c}_{k_i} = \left(\begin{array}{c} \mathbf{c}_S^{k_i} \\ 0 \end{array}\right)
\end{equation}
and the $|k_i\rangle$ are solutions to,

\begin{equation}
 \left( \mathbf{H}_0 - E \mathbf{S}_0 \right) \mathbf{c}_{k_i} = 0
\label{eq_cki}
\end{equation}

\begin{equation}
 \mathbf{H}_0=\left(\begin{array}{cc} \mathbf{H}_{SS} & 0 
      \\ 0 & \mathbf{H}_{AA} 
      \end{array}\right)
\label{eq_H0}
\end{equation}

\begin{equation}
 \mathbf{S}_0=\left(\begin{array}{cc} \mathbf{S}_{SS} & 0 
      \\ 0 & \mathbf{S}_{AA} 
      \end{array}\right)
\label{eq_S0}
\end{equation}

The matrix version of equation (\ref{eq_LS}) is,

\begin{eqnarray}
\left(\begin{array}{c} \mathbf{c}_S \\ \mathbf{c}_A \end{array}\right) = \left(\begin{array}{c} \mathbf{c}_S^{k_i} \\ 0 \end{array}\right) +  && \left(\begin{array}{cc} \mathbf{G}^r_{SS} & \mathbf{G}^r_{SA} 
      \\ \mathbf{G}^r_{AS} & \mathbf{G}^r_{AA} 
      \end{array}\right) \nonumber \\
\times && \left(\begin{array}{cc} \mathbf{V}_{SS} & \mathbf{V}_{SA} 
      \\ \mathbf{V}_{AS} & \mathbf{V}_{AA} 
      \end{array}\right) \left(\begin{array}{c} \mathbf{c}_S^{k_i} \\ 0 \end{array}\right)
\label{eq_LS_Mat}
\end{eqnarray}
and the Green's functions matrices are defined as,

\begin{equation}
 \mathbf{G}^r(E) = \left( E^+\mathbf{S} - \mathbf{H} \right)^{-1}
\end{equation}

\begin{equation}
  \mathbf{G}^{0r}(E) = \left( E^+\mathbf{S}_0 - \mathbf{H}_0 \right)^{-1}
\end{equation}

\begin{equation}
 E^+ = \lim_{\delta \to 0} E + i\delta
\end{equation}

What we need next, is to determine the perturbation matrix
$\mathbf{V}$ entering equation (\ref{eq_LS}). By multiplying equation (\ref{eq_LS_Mat}) by $\left(
E^+\mathbf{S} - \mathbf{H} \right)$ from the left, we get that
$\mathbf{V}_{SS}=0$ and $\mathbf{V}_{AS}=\mathbf{H}_{AS} - E
\mathbf{S}_{AS}$, where we have used equation (\ref{eq_cki}) and the
fact that,

\begin{equation}
 \lim_{\delta \to 0} i\delta \mathbf{S} \mathbf{c}_{k_i} = 0
\end{equation}
Using the second part of equation (\ref{eq_LS}) -- the one that
involves $\hat{G}^{0r}$ -- in a matrix form equivalent to equation
(\ref{eq_LS_Mat}), we get that $\mathbf{V}_{AA}=0$. Imposing $\mathbf{V}$
to be hermitian, we finally have,

\begin{equation}
 \mathbf{V}=\left(\begin{array}{cc} 0 & (\mathbf{H}_{SA}-E \mathbf{S}_{SA})
      \\ (\mathbf{H}_{AS}-E \mathbf{S}_{AS}) & 0 
      \end{array}\right)
\label{eq_V}
\end{equation}
and we can see that
$(\mathbf{H}_0+\mathbf{V}-E\mathbf{S}_0)\mathbf{c}_{\psi}=0$. The
energy-dependent form of $\mathbf{V}$ can be traced back to the the use
of a non-orthogonal basis set~\cite{Lang_GF,Emberly_NonOrt}.

The value of $\langle k_i|\hat{T}| k_f\rangle$ can thus be calculated with,

\begin{eqnarray}
\langle k_i|\hat{T}| k_f\rangle &=& \left(\begin{array}{cc} \mathbf{c}^{k_i*}_S & 0 \end{array}\right).\left(\begin{array}{cc} \mathbf{T}_{SS} & \mathbf{T}_{SA}
      \\ \mathbf{T}_{AS} & \mathbf{T}_{AA} 
      \end{array}\right).\left(\begin{array}{c} \mathbf{c}^{k_f}_S \\ 0 \end{array}\right)
\label{eq_T_exp_val} \nonumber \\
&=& \mathbf{c}^{k_i*}_S\mathbf{T}_{SS}\mathbf{c}^{k_f}_S \nonumber \\
&=& \mathbf{c}^{k_i*}_S\left(\mathbf{V}_{SA}\mathbf{G}^r_{AA}\mathbf{V}_{AS}\right)\mathbf{c}^{k_f}_S
\end{eqnarray}
where we have used equation (\ref{eq_T_op}). Substituting this back into
equation (\ref{eq_T_ST}), we get that,

\begin{widetext}
\begin{eqnarray}
 \frac{T(E_F)}{(2\pi)^2} &=& \sum_{k_i,k_f}  \mathbf{c}^{k_i*}_S\left(\mathbf{V}_{SA}\mathbf{G}^r_{AA}\mathbf{V}_{AS}\right)\mathbf{c}^{k_f}_S   
                                 \mathbf{c}^{k_f*}_S\left(\mathbf{V}_{SA}\mathbf{G}^a_{AA}\mathbf{V}_{AS}\right)\mathbf{c}^{k_i}_S
\delta(\varepsilon_i-\varepsilon_f)\delta(\varepsilon_i-E_F),\nonumber \\
&=& Tr\left[  \mathbf{V}_{AS}\left(\sum_{k_i} \mathbf{c}^{k_i}_S\mathbf{c}^{k_i*}_S \delta(\varepsilon_i-E_F) \right) 
               \mathbf{V}_{SA}\mathbf{G}^r_{AA}\mathbf{V}_{AS}
                \left(\sum_{k_f} \mathbf{c}^{k_f}_S\mathbf{c}^{k_f*}_S \delta(\varepsilon_i-\varepsilon_f) \right)
                \mathbf{V}_{SA}\mathbf{G}^a_{AA}\right] \nonumber \\
&=& \frac{1}{(2\pi)^2}Tr\left[ \mathbf{\Gamma}_{AA}(E_F)\mathbf{G}^r_{AA}(E_F)\mathbf{\Gamma}_{AA}(\varepsilon_i=E_F)\mathbf{G}^a_{AA}(E_F)\right]
\label{eq_T_ST_final}
\end{eqnarray}
and we obtain the final expression for $T(E_F)$,

\begin{equation}
 T(E_F) = Tr\left[ \mathbf{\Gamma}_{AA}(E_F)\mathbf{G}^r_{AA}(E_F)\mathbf{\Gamma}_{AA}(E_F)\mathbf{G}^a_{AA}(E_F)\right]
\label{eq_T_E_final}
\end{equation}

\end{widetext}
In the derivation of equation (\ref{eq_T_ST_final}) we have used that:

\begin{enumerate}
 \item $\mathbf{c}^{j}_S\mathbf{c}^{j*}_S$ can be considered to define a
matrix: $\mathbf{c}^{j}_S$ being a column vector, and $\mathbf{c}^{j*}_S$
a row vector.
 \item The cyclic properties when taking the trace of
 a product of matrices.  
\item The discontinuity of the \textit{retarded} and \textit{advanced} Green's functions at the real axis~\cite{Economou} that in our basis set, considering item 1, gives,
\begin{equation} \sum_{j}
 \mathbf{c}^{j}_S\mathbf{c}^{j*}_S \delta(E-\varepsilon_j) =
 \frac{i}{2\pi}(\mathbf{G}^{0r}_{SS}(E)-\mathbf{G}^{0a}_{SS}(E))
 \end{equation} 
\item The \textit{self-energy} and \textit{gamma} matrices are written as~\cite{Transiesta},
\begin{equation}
  \mathbf{V}_{AS} \mathbf{G}^{0r(a)}_{SS} \mathbf{V}_{SA}  =
\mathbf{\Sigma}^{r(a)}_{AA} \label{eq_self_en}
  \end{equation} 
 \begin{equation}
   \mathbf{\Gamma}_{AA} = i \left[ \mathbf{\Sigma}^{r}_{AA} -
\mathbf{\Sigma}^{a}_{AA} \right] \label{eq_gamma}
 \end{equation}

\end{enumerate}

We now describe the procedure to obtain $T(E_F)$, defined in equation
(\ref{eq_T_E_final}), from our ab-initio simulations. The ideas are
the same as the ones used to derive the equations for the transmission
function for electronic transport calculations, but since the final
formulas are not exactly the same, we here include the derivation of
the relations that we have used in the present work. To do so, we start
by re-writing the DFT Hamiltonian (equation (\ref{eq_DFT_Mat})),

\begin{equation}
 \mathbf{H}=\left(\begin{array}{ccc} \mathbf{H}_{EE} & \mathbf{H}_{EC} & 0
      \\ \mathbf{H}_{CE} & \mathbf{H}_{CC} & \mathbf{H}_{CA} \\
      0 & \mathbf{H}_{AC} & \mathbf{H}_{AA}
      \end{array}\right)
\label{eq_DFT_Mat_ext}
\end{equation} 
where $\mathbf{H}_{EE}$ is a semi-infinite matrix describing the
semi-infinite electrode (a region  where the electronic structure,
and matrix elements are assumed to be already bulk-like); and
$\mathbf{H}_{CC}$ is the ``slab''  that describes the actual surface. The
size of the $C$ (for \textit{contact}) region is in principle arbitrary
(but finite!), as long as it is thick enough to have: \textit{i)} the
$\mathbf{H}_{EA}$ ($\mathbf{H}_{AE}$) matrix elements equal to zero;
and \textit{ii)} the $\mathbf{H}_{EE}$ matrix elements sufficiently
converged to bulk values.

For the purpose of simplifying the notation, let us define an $\mathbf{h}$ matrix to be,

\begin{equation}
 \mathbf{h} = E\mathbf{S} - \mathbf{H}
\end{equation}
To compute $T(E_F)$, we need $\mathbf{G}^r_{AA}(E_F)$ and $\mathbf{\Gamma}_{AA}(E_F)$, where,

\begin{equation}
\mathbf{G}^r_{AA} = \left( \mathbf{h}_{AA} - \mathbf{\Sigma}_{AA} \right) 
\label{eq_gf_at}
\end{equation}
From Eqs.~(\ref{eq_self_en}), (\ref{eq_gamma}) and (\ref{eq_gf_at}),
we can see that what we need is to compute $\mathbf{G}^{0r(a)}_{CC}$,
the finite portion of $\mathbf{G}^{0r(a)}_{SS} $ needed to compute the
self energies,

\begin{equation}
  \mathbf{G}^{0r}_{CC} = \left( \mathbf{h}_{CC} -\mathbf{h}_{CE}\mathbf{G}^{0r}_{EE}\mathbf{h}_{EC} \right)
\end{equation}
where it is important to note that only a finite number of elements of
$\mathbf{h}_{CE}$ ($\mathbf{h}_{EC}$) have non-zero values, hence only a
finite portion of $\mathbf{G}^{0r}_{EE}$ needs to be calculated. Since the
$\mathbf{h}_{EE}$ matrix elements are assumed to be already ``bulk-like'',
the $\mathbf{G}^{0r}_{EE}$ matrix is obtained using the matrices extracted
from a bulk calculation of the electrode\cite{Lopez_Algo,Transiesta}.

From these equations, we see that $T(E_F)$ represents the flux of
electrons coming from the substrate and scattering off the adsorbate
back into the substrate. This quantity is different from the transmission
function appearing in a Landauer-like approach~\cite{Transiesta}, where
the adsorbate is connected to two electrodes. The difference is clear in
the above equations, here a unique reservoir (the substrate) appears in
the self-energies, and hence the decay appearing in the elastic Green's
function, Eq.~(\ref{eq_gf_at}), has only one self-energy instead of two
in the transport case~\cite{Transiesta}.

\section{Density-functional study of a single M\lowercase{n} adsorbate on a C\lowercase{u}N/C\lowercase{u}(100) surface}
\label{DFT_Results}

\begin{figure}
\includegraphics[angle=0,width=0.35\textwidth]{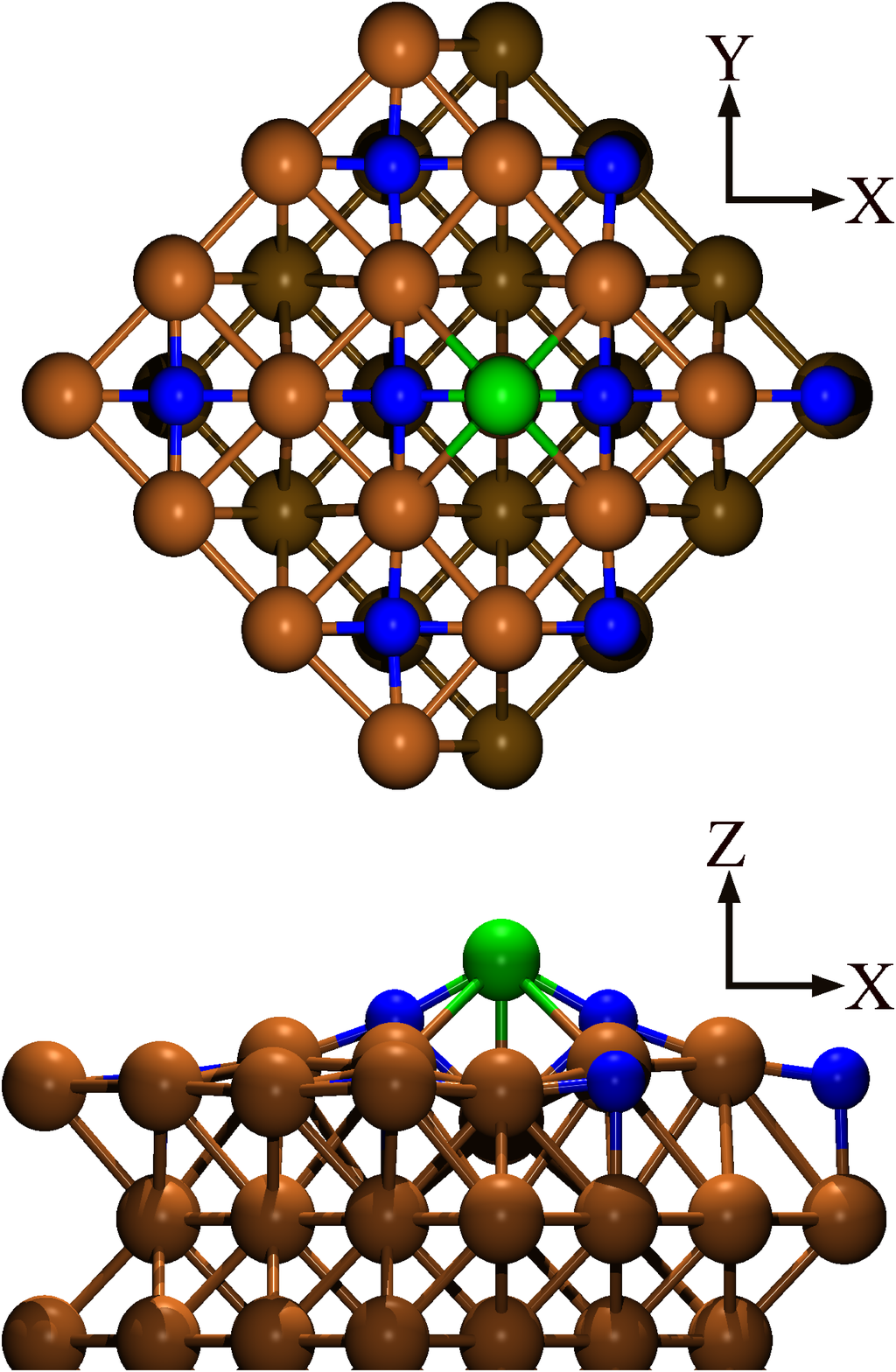}
\caption{
Top and lateral view of the geometry used to describe the Mn adsorbate
(green) on top of a CuN/Cu(100) surface. The Mn atom is located on top
of a Cu atom of the CuN layer, pushing it down while pulling up the N
(blue) atoms close to it.  } \label{fig_geom_Mn_CuN} \end{figure}

The ground-state electronic-structure configuration and the value
for $T^{Total}(E_F)$ were obtained by density-functional-theory
(DFT) simulations.  Our DFT calculations were performed using the
SIESTA code\cite{Siesta}. The super cell contained at least six $4\times4$
Cu (100) layers -- a larger number of layers was used to test the
convergence with respect to the size of the $C$ region, as discussed
in section \ref{DFT-Model}; one CuN layer; and one Mn atom (see
Fig. \ref{fig_geom_Mn_CuN}). The Mn atom and the two outermost layers
were relaxed until the forces were smaller than 0.03 eV/Ang. A sampling
of $3\times3\times1$ k-points was used. We have used the generalized gradient
approximation~\cite{PBE} for the exchange-correlation potential.

\begin{figure}
\includegraphics[angle=0,width=0.45\textwidth]{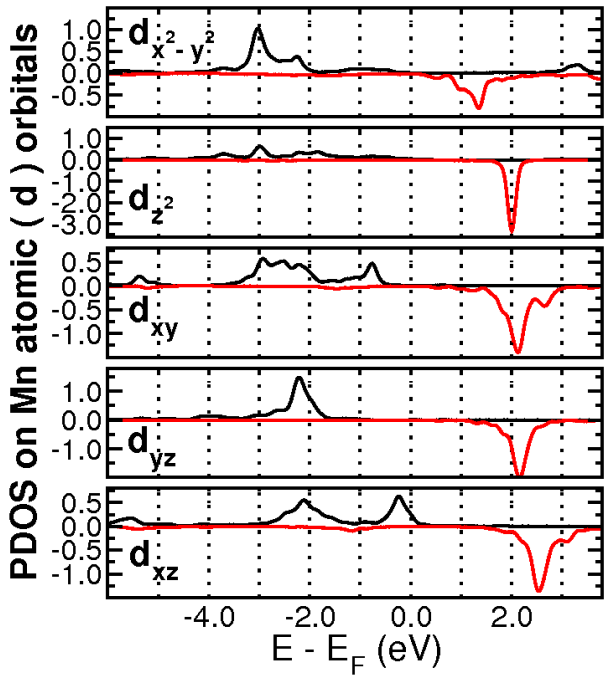}
\caption{
Projected density of states (PDOS) on the Mn atomic orbital for the case of Mn on CuN/Cu(100). For all
the curves shown here, the positive (black) curves corresponds to the
 majority spin, and the negative (red) to the minority spin.
}
\label{fig_PDOS_Mn_CuN}
\end{figure}

We have obtained a total spin polarization ($Q_{up}-Q_{down}$) of 4.7 Bohr magnetons,
which essentially corresponds to a $S=\frac{5}{2}$ spin configuration,
in good agreement with experiment~\cite{Loth10}. The spin polarization
is localized around the Mn atom, that has five half-filled d-orbitals,
corresponding to five unpaired electrons. This can clearly be seen in
Fig.~\ref{fig_PDOS_Mn_CuN}.

We now turn to the calculation of $T^{Total}(E_F)$. As already discussed
in section \ref{DFT-Model}, the thickness of the $C$ region of Eq.~(\ref{eq_DFT_Mat_ext}) 
is arbitrary (to some extent), and here we have taken
the first three surface layers (including the CuN layer), as shown in
Fig. \ref{fig_geom_Mn_CuN}.  Since there is also some arbitrariness in
the definition of what \textit{the atom is}, we have considered different
basis sets: from SZP to DZP, with larger and shorter cutoff radii. The
differences in the results were not appreciable, and so here we report
just the results for a DZP basis with large cutoff values --- up to 10
~\AA~ for the Mn basis orbitals.

\begin{figure}
\includegraphics[angle=0,width=0.45\textwidth]{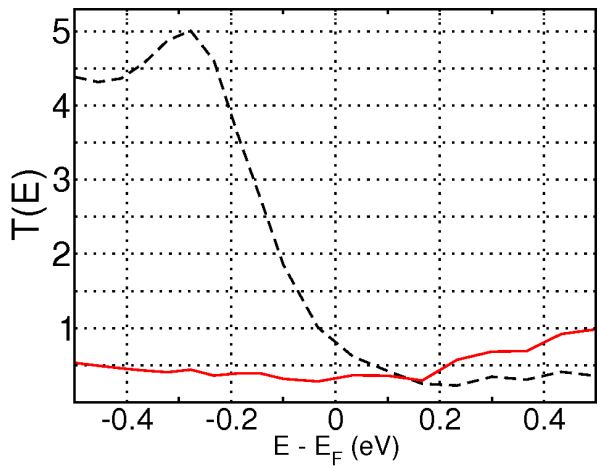}
\caption{
T(E) for the case of Mn on CuN/Cu(100). The black (dashed) curve
 corresponds to the majority spin, and the red (continuous) to the
 minority spin. $T^{Total}(E_F)$ is the sum
of T(E) for both spins at $E=E_F$. A sampling of $15\times15$ k-points was
used to obtain converged values for T(E).  } \label{fig_T_Mn_CuN}
\end{figure}

We have computed T(E) for a range of energy values
around the Fermi level, and the results are show in
Fig.~\ref{fig_T_Mn_CuN}. As for $T^{Total}(E_F)$, we have found
$T^{Total}(E_F)=T(E_F,\uparrow)+T(E_F,\downarrow)=0.8+0.3=1.1$.
When we analyze the orbital contribution to
 the function T(E) of Fig.~\ref{fig_T_Mn_CuN}, we identify the
d$_{xz}$ orbital of Mn as responsible for the electron scattering off
the Mn adsorbate. 
The above information on the characteristics of the substrate electrons
hitting the adsorbate can be used to further specify the inputs of our
spin-transition calculations.  Indeed, the ground state of the system
corresponds to putting one electron into each
 Mn d orbital (defined with the appropriate symmetry). This generates
 the $S=5/2$ $M=5/2$ state of the adsorbate.  The dominant contribution
 to substrate electrons going through the adsorbate is found to
 involve the $d_{xz}$ orbital with majority spin as the transition
 intermediate; this is interpreted as a  process involving a positive
 ion intermediate of $S_T = 2$ symmetry. Similarly, the contribution
 associated to substrate electrons going through the $d_{xz}$ orbital
 with minority spin is interpreted as a process involving a negative ion
 intermediate of $S_T = 2$ symmetry. So in all cases, the deexcitation
 process induced by substrate electrons going through the Mn adsorbate
 involves a $S_T = 2$ intermediate and the associated electron flux is
 given by $T^{Total}(E_F)= 1.1$.

At this point, one can stress the stark contrast between
the present deexcitation study and  our earlier study on magnetic
excitation by electron tunnelling between the tip and the substrate
(Ref.~[\onlinecite{Lorente09}]). In the tunnelling electron case,
 the Mn orbitals contributing to the transmission were the extended
$s$ and $p$ orbitals whereas here, for the electrons scattering from
the substrate into the substrate via the adsorbate, a $d$ orbital  is
dominating. In addition, the spin symmetry of the scattering intermediates
are different : $S_T = 2$ vs $S_T = 3$. This study permits us to conclude
that the de-excitation of spin states via electron-hole pairs takes place
through the Mn $d$ electrons and in particular the d$_{xz}$ orbital,
while the spin excitation process proceeds via the tunneling electrons
and are of $s$ and $p$ characters.

\section{Lifetime of the excited magnetic states}
\label{Lifetime_mag_states}

\begin{figure}
\includegraphics[angle=0,width=0.4\textwidth]{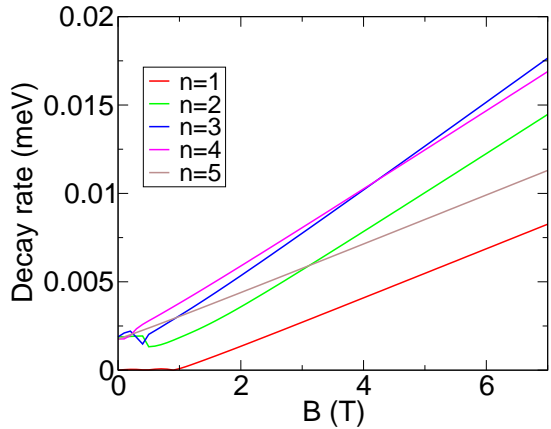}
\caption{
Decay rate (in meV) of the five excited magnetic states in the
Mn/CuN/Cu(100) system as a function of the applied magnetic field B. The
B field is along the x-axis. The decay rate is the inverse of the excited
state lifetime.  } \label{fig_decay_rates} \end{figure}

Equation (\ref{eq_decay_final}) together with the results of section
\ref{DFT_Results} has been used to compute the decay rates of the
excited states of the system. Figure \ref{fig_decay_rates} presents
the total decay rate, $\Gamma_{Tot,i}$, inverse of the lifetime, of
the five excited states as a function of the applied magnetic field
B. Two different regimes can be seen:  low B and large B. At large B,
the magnetic structure of the system is quasi-Zeeman. As a consequence,
the decay of an excited state is dominated by one channel corresponding
to a $\Delta M_x = -1$ selection rule. The B dependence of the decay
rate is then that of the energy change associated to the decay, i.e. it
is linear in B with a slope proportional to the dominant spin transition
probability, $P_{Spin}$. At small B, the variation is more complex,
reflecting the complex decoupling of the anisotropy by the B field on
the x-axis (see Fig.~\ref{fig_Mag_En_Lev}). However, one can notice
that the lowest excited state remains quasi-degenerate with the ground
state almost up to 1T (Fig.~\ref{fig_Mag_En_Lev}), so that its decay
rate is extremely small due to a quasi-vanishing $\delta\Omega_f$ (see
Eq. (\ref{eq_decay_final})). Actually, this simply means that at low
B, for a finite temperature, the two lowest states are roughly equally
populated. As for the states 2-5, at low B, their decay rate is in the
2.0 $\mu eV$ range, corresponding to a lifetime of the order of 0.3 ns.

One can stress that a different direction of the B-field leads to a
different behaviour of the decay rates. The limit B = 0 is the same,
obviously and the large B is almost independent of the B direction in
the present case, but not completely due to an incomplete decoupling of
the magnetic anisotropy for B = 7 T. The decay rates in the intermediate
region, where the decoupling of the anisotropy occurs, depend of the
direction of the B field with respect to the magnetic axis of the system.

\begin{figure}
\includegraphics[angle=0,width=0.4\textwidth]{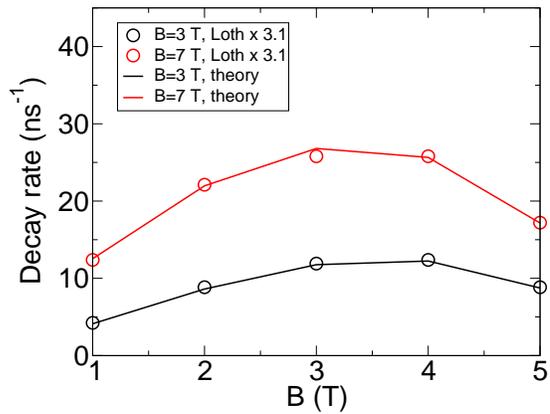}
\caption{
Comparison between the present decay rate (full lines) of the five excited
magnetic states in the Mn/CuN/Cu(100) system with the experimental
results of Loth et al~\cite{Loth10} (symbols). The black symbols and
line correspond to an applied B-field of 7T and the red symbols and line
to 3T (the B-field is along the x-axis).  For the sake of comparison,
the experimental results of Loth et al have been multiplied by a global
factor equal to 3.1.} \label{fig_decay_rates_comp} \end{figure}

Figure \ref{fig_decay_rates_comp} presents a direct comparison between
the present decay rate of the five excited states of Mn on CuN/Cu(100)
and the decay rate extracted by Loth et al~\cite{Loth10} from their
experimental data. Two values of the magnetic field are presented: 3 and
7 T. For the sake of comparison, the experimental results of Loth et al
have been multiplied by a global factor equal to 3.1.  It appears that
the present study reproduces extremely well the state dependence and
the magnetic field dependence of the excited state lifetimes. However,
the present results for the decay rates are a factor 3 larger than the
experimentally extracted data. Besides inaccuracies and approximations
in the experimental and theoretical procedures, one can invoke the
sensitivity of the present results on the energy position of the $d_{xz}$
orbital in the calculations. Indeed, as we can see in Fig. 3 and 4, the
electron flux hitting the adsorbate at Fermi level corresponds to the tail
of the $d_{xz}$ majority and minority spin orbitals and any inaccuracy
in the orbital energy directly affects, $T(E_F)$, the electron flux. 

\section{Modelling of the M\lowercase{n} junction conductance at large current}

The experiments of Loth et al~\cite{Loth10} introduce two main effects
compared to the earlier ones in Ref.[\onlinecite{Hirjibehedin06}]. First,
by using a polarized tip, these experiments introduce an unbalance
between the spin up and spin down tunneling electrons that reveals the
spin-dependence of the junction conductance. Second, they consider large
currents flowing through the junction allowing a stationary population
of excited states induced by the tunneling electrons. Indeed the two
effects are linked together, tunneling electrons of different spins
leading to different excited state populations.  Below, we examine these
two effects in the Mn on CuN/Cu(100) system allowing a modelling of the
experimental situation.

\subsection{Spin-dependence of the conductance of the adsorbed Mn atom}
\label{spin_dep_cond}

In references [\onlinecite{Lorente09,Gauyacq10}], we developed a
strong coupling treatment of the elastic and inelastic tunneling
of electrons through a magnetic adsorbate that inspired the above
treatment of excited magnetic state decay. As explained in section
~\ref{Method}, the magnetic anisotropy terms are small, and one can
treat them in the sudden approximation and define a tunneling amplitude,
independent of the magnetic anisotropy. As the main result (see details in
[\onlinecite{Gauyacq10}]), the system conductance for the Mn atom in the
magnetic state i, (in the case of only one $S_T$ symmetry contributing
to tunneling) is given by:

\begin{equation}
 \frac{dI}{dV}=C_0 \frac{\sum_n\Theta(V-E_n)\sum_{m,m'}\left|\sum_j A_{j,i,m}A^*_{j,n,m'}\right|^2}{\sum_n\sum_{m,m'}\left|\sum_j 
A_{j,i,m}A^*_{j,n,m'}\right|^2}
\label{eq_diff_cond}
\end{equation}
where $C_0$ is a global conductance; on the small energy range that
we consider here and for a fixed tip-adsorbate distance, $C_0$ can be
considered as constant. V is the junction bias and $E_n$ the energy of the
various magnetic states, n, of the system. The $A_{j,n,m}$ coefficients
are spin coupling coefficients giving the expansion of the eigenstates
of $\vec{S}_T^2$ and $S_{T,z}$ on the initial or final states of the
tunneling process, $|m,\phi_n\rangle$, (m refers to the tunneling electron
spin state and $\phi_n$ is the magnetic anisotropy state of the adsorbate,
eigenstate of the Hamiltonian (\ref{hamiltonien})):

\begin{equation}
 A_{j,n,m}=\langle S_T,M_T|m,\phi_n\rangle \textit{,  with } j=(S_T,M_T)
\end{equation}
In both treatments (inelastic tunneling and excited state decay),
the final result appears as a product of a magnetism-free quantity
by a spin-coupling coefficient term, i.e. a global conductivity of
the system is shared among the various anisotropy channels. The
excitation probability is then only dependent on the weight of
the incident and final channels in the intermediate tunneling
intermediate and it can be very large. Such an efficient inelastic
process has been invoked in several other processes involving angular
momentum transfer (rotational or spin) in gas phase or surface
problems~\cite{Abram69,Teillet-Billy87,Bahrim94,Teillet-Billy00},
in all cases, it lead to high probabilities of inelastic scattering
(see also a discussion in Ref. [\onlinecite{Gauyacq10}]). A change
in the adsorbate-tip distance leads to a change in $C_0$ only and
consequently in the tunneling current, but without a change in the
excitation probabilities.

\begin{figure}
\includegraphics[angle=0,width=0.4\textwidth]{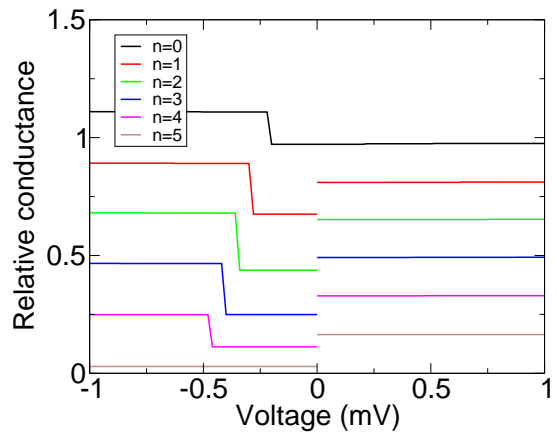}
\caption{
Conductance of the various magnetic states in the Mn/CuN/Cu(100) system
as a function of the tip bias in mV for a fully polarized electrode
(spin down). The B-field is along the x-axis and is equal to 3T. The
conductance has been normalized in such a way that the conductance for
the ground state and a non-polarized tip is equal to 1 at zero bias.
} \label{fig_relat_cond_down} \end{figure}

\begin{figure}
\includegraphics[angle=0,width=0.4\textwidth]{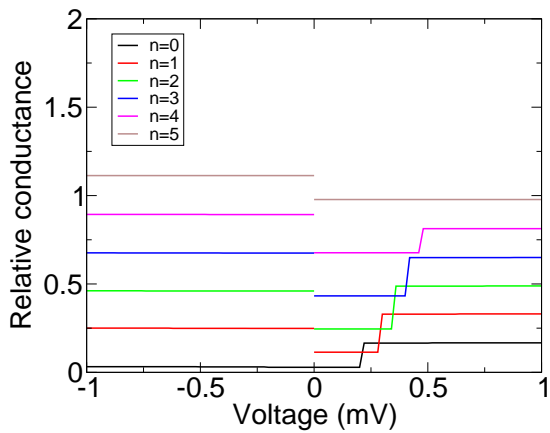}
\caption{
Conductance of the various magnetic states in the Mn/CuN/Cu(100) system
as a function of the tip bias in mV for a fully polarized electrode
(spin up). The B-field is along the x-axis and is equal to 3T. The
conductance has been normalized in such a way that the conductance for
the ground state and a non-polarized tip is equal to 1 at zero bias.
} \label{fig_relat_cond_up} \end{figure}

\begin{figure}
\includegraphics[angle=0,width=0.4\textwidth]{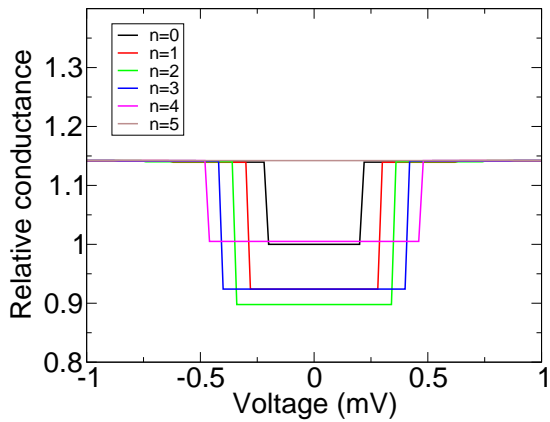}
\caption{
Conductance of the various magnetic states in the Mn/CuN/Cu(100)
system as a function of the tip bias in mV for a non-polarized
electrode. The B-field is along the x-axis and is equal to 3T. The
conductance has been normalized in such a way that the conductance for
the ground state and a non-polarized tip is equal to 1 at zero bias.
} \label{fig_relat_cond_non_pol} \end{figure}

In the case of Mn on CuN/Cu(100), a DFT calculation showed that the
$S_T$ = 3 symmetry is dominating the tunneling process~\cite{Lorente09}
and this accounted well for the observations of Hirjibehedin et
al~\cite{Hirjibehedin06}, obtained with non-polarised tunneling
electrons. 
In Ref. [\onlinecite{Lorente09,Gauyacq10}], we only considered tunneling
of non-polarised electrons, i.e. we summed the contributions from the
two electron spin directions, both in the incident and final channels
(the sum over $m$ and $m'$ in equation (\ref{eq_diff_cond})). Here we
consider tunneling for a fixed direction of the electron spin (fixed
$m$ or $m'$) in the incident or final state (the spin directions are
defined along the x-axis parallel to the applied B field). We thus use
equation (\ref{eq_diff_cond}) with the sum over $m$ (or the sum over
$m'$) removed from the numerator. Figures \ref{fig_relat_cond_down}
and \ref{fig_relat_cond_up} present the relative conductance of the
various magnetic states of Mn for a fully polarized electrode and a B
field of 3 T. Figure \ref{fig_relat_cond_up}  corresponds for $V > 0$ to
an incident electron with an 'up' spin and for $V<0$ to an electron in
a final 'up' state. Fig. \ref{fig_relat_cond_down} presents the 'down'
equivalent. In both cases the conductance is normalised in such a way
that the conductance at $V= 0$ for a non-polarised beam is equal to 1. The
conductance presents steps at the magnetic excitation thresholds, due to
the excitation induced by the tunneling electrons; the conductance also
takes into account the possibility of de-excitation processes induced
by the tunneling electrons, these present no energy thresholds. No
broadening effect has been introduced in the conductance in Figs.
\ref{fig_relat_cond_down},\ref{fig_relat_cond_up},\ref{fig_relat_cond_non_pol}
and the inelastic steps should be vertical; the finite slope visible in
the figures comes from the finite number of V points that were actually
computed.

\begin{figure*}
\includegraphics[angle=0,width=0.80\textwidth]{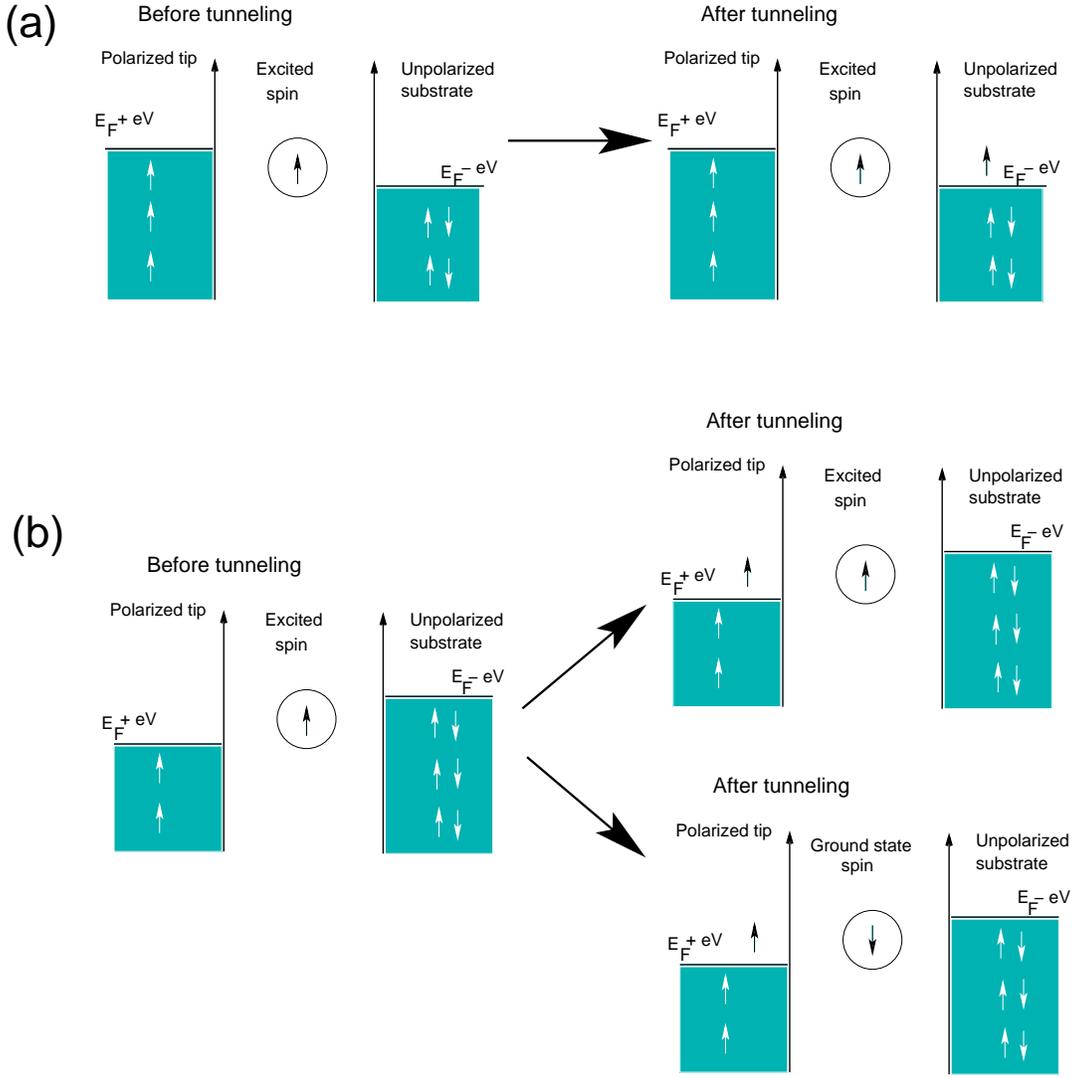}
\caption{Qualitative scheme of the bias asymmetry in the tunneling process.
For clarity of the illustration, we assume a $S=\frac{1}{2}$ adatom in its 
`spin-up' state. The tip is completely polarized along the `up' direction.
(a) For positive bias, $V>0$, electron tunneling can only be elastic because
the tip is spin polarized and aligned with the adatom's spin. (b) For negative
bias, $V\leq 0$, unpolarized electrons flow from the substrate and they can
tunnel either elastically or inelastically. Hence, tunneling is 
different in cases (a) and (b) and so it presents a bias asymmetry due
to the spin polarization of the tip.
}
\label{Schema}
\end{figure*}

It appears that the conductances of the various anisotropy states for the
two spin directions are quite different both in magnitude and shape (some
have inelastic contributions and some do not). This is not surprising within
our strong coupling approach: the magnitude of the conductivity
is directly given by the weights of the initial and final channels
(the $A_{j,n,m}$ coefficients)  in the tunneling symmetry ($S_T = 3$
in the present case) and these are strongly dependent on the considered
initial and final states. One can also notice that the conductivity
shown in figures \ref{fig_relat_cond_down} and \ref{fig_relat_cond_up}
are discontinuous at $V = 0$ (except for the ground state) due to
the switch of definition at $V = 0$ (spin selection for the initial
state vs spin selection for the final state). The behaviour seen in
\ref{fig_relat_cond_down} and \ref{fig_relat_cond_up} can be easily
understood for the 'large' B case depicted there. In that case, the
anisotropy states are roughly Zeeman states, eigenstates of $S_x$
with the eigenvalues $M_x$. In that case, the increasing order of
energy of the magnetic states corresponds to the increasing order
of $M_x$ and the transitions induced by a tunneling electron verify
the $\Delta M_x=\pm 1$ selection rule (the rule is strict only in the
perfect Zeeman limit). As a consequence, excitation corresponds to a
$\Delta M_x=+ 1$ selection rule and it can only exist for an incident
'up' electron and an outgoing ‘down’ electron, and this appears
clearly in Fig. \ref{fig_relat_cond_down} and \ref{fig_relat_cond_up},
where the conductivity exhibits an inelastic step for only one sign
of V, different for spins up and down. The inelastic steps appear as
different energies for the different excited states, this is due to the
fact that at 3 T the structure is not yet a perfect Zeeman structure (see
Fig. \ref{fig_Mag_En_Lev}), in this limit all inelastic steps would be at
the same position given by $g\mu_BB$. Similarly, the very small excitation
steps appearing at higher energy in addition to the $\Delta M_x=+1$
selection rule steps, as well as those appearing in the 'forbidden' $V$
side,  are due to the small difference from a pure Zeeman structure. The
discontinuity in the conductance at $V = 0$ is due to the existence of
de-excitation processes induced by the tunneling electrons. Similarly to
the excitation processes (well visible in Fig. \ref{fig_relat_cond_down}
and \ref{fig_relat_cond_up}), these are highly dependent on the sign of V,
leading to a discontinuity at $V = 0$. The purely elastic conductance
is continuous at V = 0, as is the ground state conductance. Figure~\ref{Schema}
shows a qualitative picture explaining this bias asymmetry.

Figure \ref{fig_relat_cond_non_pol} presents the relative conductance
for the various magnetic anisotropy states in the case of a
non-polarised tip. Similarly to Fig. \ref{fig_relat_cond_down} and
\ref{fig_relat_cond_up}, the conductance has been normalised so that
the conductance at $V = 0$ for the ground state is equal to 1. All
the conductivities are now continuous at $V = 0$ and symmetric in V
since no selection rule is imposed on the tunneling electron spin. The
conductivity for each magnetic state exhibits a large inelastic step
(except the highest lying state!) corresponding to the $\Delta M_x=\pm
1$ selection rule; much smaller excitation steps also appear at higher
energies (barely visible in the figure) due to the non-perfect Zeeman
structure. For all excited states, inelastic tunneling is significant
compared to elastic tunneling, it is in the 15-25 \% range, and different
for the different states. One can also notice in Fig. \ref{fig_relat_cond_non_pol} that the conductivities at
large bias for all the states are equal, as a consequence of the summation
over all possible excited states in Eq. \ref{eq_diff_cond}. One can then
conclude from Figs. \ref{fig_relat_cond_down}-\ref{fig_relat_cond_non_pol}
that the conductivity in the case of a polarised tunneling electron is
quite different from that for a non-polarised one; in general,  simply
looking at majority and minority spins is insufficient, one has to resort
to spin coupling arguments.

\subsection{Description of the conductivity in the presence of excited states}

The excitation of magnetic states by tunneling electrons is very
efficient  in the Mn/CuN/Cu(100). Though not as efficient as in some
metal-phthalocynanine case~\cite{Tsukahara,XiChen,Gauyacq10},  this
can lead to a significant stationary population of excited states in
an experiment with a finite current. This effect has been very clearly
demonstrated by Loth et al~\cite{Loth10}. We modelled this situation in
a way very similar to that used by Loth et al~\cite{Loth10}. The main
difference lies in the absence of adjustment parameters in the present
work: with the present treatment of the excitation process (section
\ref{spin_dep_cond}) and of the decay process (sections \ref{Method} and
\ref{Lifetime_mag_states}), we can quantitatively predict the behaviour
of the conductance as a function of the tunneling current.

The basic idea is to compute the stationary population of excited states
that is induced by the tunneling current and then, via the excited
conductance discussed in \ref{spin_dep_cond}, to get the junction
conductivity. For an STM tip positioned above the Mn adsorbate, a tip
bias V and a tunneling current I, the time dependence of the population,
$P_i(I,V)$, of the magnetic state, i, is given by:

\begin{eqnarray}
 \frac{dP_i(I,V)}{dt}=&-&P_i(I,V)\left(\sum_j \Gamma_{i,j} + \sum_j F_{i,j}(I,V) \right) \nonumber \\
                      &+& \sum_j P_j(I,V) \left( F_{j,i}(I,V)+\Gamma_{j,i}\right)
\label{eq_master}
\end{eqnarray}
where $\Gamma_{i,j}$ is the partial decay rate of state i towards state j
(sections \ref{Method} and \ref{Lifetime_mag_states}).  $F_{i,j}(I,V)$  is
 the transition rate from state i to state j induced by the tunneling
electrons (section \ref{spin_dep_cond}). It is given by:

\begin{equation}
 F_{i,j} = C P_{Spin}(i,j)\Delta_{i,j}(V)
\end{equation}
where $P_{Spin}(i,j)$ is the spin coupling coefficient of the considered
$i \rightarrow j$ transition (section \ref{spin_dep_cond}) for the
considered spin states of the tunneling electron. $\Delta_{i,j}(V)$
is an energetic factor; for the vanishing temperature considered here,
it is equal to $V+E_i - E_j$ for an open excitation channel ($i
\rightarrow j$ transition), to 0 for a closed excitation channel and
to V for a de-excitation channel. C is a factor corresponding to the
global conductance of the system. Below, the factor C is set so that
the junction conductance at $V = 0$, G, has a fixed value and then the
whole spectrum of conductivity as a function of V is computed; changing
C corresponds to moving the tip with respect to the Mn adsorbate i.e. to
changing the magnitude of the tunneling current.

Equation (\ref{eq_master}) yields the time dependent population of
the excited states, including the transient regime at the switching of
the applied bias. The experiments being slow on the time scale of the
excited state relaxation time (typically a fraction of ns, as seen in
Fig. \ref{fig_decay_rates}), the populations quickly reach stationary
values which determine the observed conductivity. These stationary values
are obtained by solving the homogeneous set of equations obtained from
(\ref{eq_master}) by setting all time derivatives to zero. Once the
populations are known the junction conductance is obtained by summing
the contributions of the different states.

\subsection{Population of the excited states}

Loth et al~\cite{Loth10} performed their experiments with a partial
polarisation of the tip typically equal to $\eta=0.24$ so that the
two spin directions of the electron have probabilities equal to 0.5
($1\pm\eta$). In the present system, the ground state at large B is almost
the $M_x=-\frac{5}{2}$ state. The polarisation of the tip is in the same
direction, so that electrons with spin down are dominating at $V > 0$
and holes with spin down at $V < 0$.

\begin{figure}
\includegraphics[angle=0,width=0.4\textwidth]{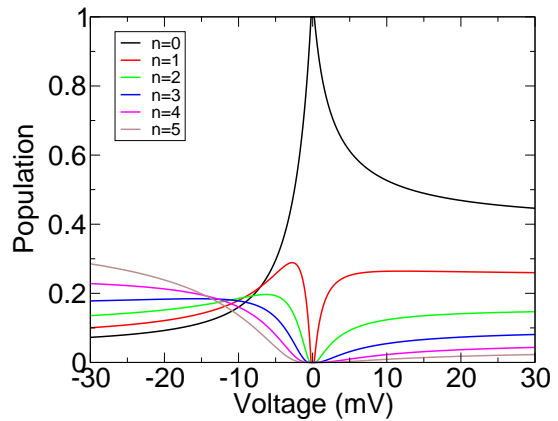}
\caption{
Population of the six magnetic states in the Mn/CuN/Cu(100) system as a
function of the tip bias. The magnetic field is along the x-axis and equal
to 3T. The tip polarisation, $\eta$, is equal to 0.24. The conductance
at zero bias is equal to $2. 10^{-6}$ S.  } \label{fig_populations_pol}
\end{figure}

The corresponding population of excited states for a junction
conductance at $V = 0$ equal to $2. 10^{-6}$ S, a magnetic B field
of 3 T and a tip  polarisation equal to $\eta=0.24$ are shown in
Fig.~\ref{fig_populations_pol} as a function of V. Below the threshold
for the $0 \rightarrow 1$ transition, the population is entirely in the
ground state, beyond this threshold the excited state population quickly
increases as $|V|$ increases. One can notice that the excitation process
at 3T basically goes step by step from state 0 to 5 following the energy
(and index) order, so that excitation of the higher lying states is
a multiple order process and rises much more slowly than excitation to
state 1. At large $|V|$, the system reaches a regime where the population
is independent of V, it corresponds to the situation where transitions
(excitation and de-excitation) induced by tunneling electrons dominate
over the excited state decay.  Excitation by electrons and holes are very
different leading to very different excited state populations 
in the $V>0 $ and $V<0$ cases. 'Down'
polarised holes have a much higher excitation efficiency than 'down'
polarised electrons, and this is a direct consequence of the effects
observed in Fig. \ref{fig_relat_cond_down} and \ref{fig_relat_cond_up}
and discussed above. Actually, if we forget the very tiny excitations
that correspond to the non-perfect Zeeman limit, a fully polarised
tip would only lead to an excited state population for $V < 0$ (see
Fig. \ref{fig_relat_cond_down}); here with a finite $\eta$ polarisation,
the excited state populations at $V > 0$ are due to the minority
spin direction electrons in the tip. Indeed for a non-polarized tip,
excitation by electrons and holes are equivalent, consistently with
Fig.~\ref{fig_relat_cond_non_pol}.

\subsection{Modelling of Loth et al experiment}

\begin{figure}
\includegraphics[angle=0,width=0.4\textwidth]{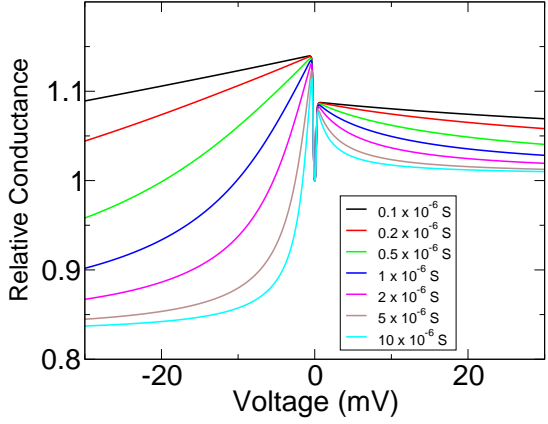}
\caption{
Relative conductance of the Mn/CuN/Cu(100) system as a function the tip
bias. The tip polarization is $\eta = 0.24$ and the B field, equal to
3T, is along the x-axis. The finite population of the excited states is
taken into account. The various curves correspond to various absolute
conductances at zero bias (0.1, 0.2, 0.5, 1., 2., 5. and 10. $10^{-6}$
S). In the figure, the conductance is plotted in relative value, with the
conductance for zero bias set to 1.  } \label{fig_cond_3T} \end{figure}

\begin{figure}
\includegraphics[angle=0,width=0.4\textwidth]{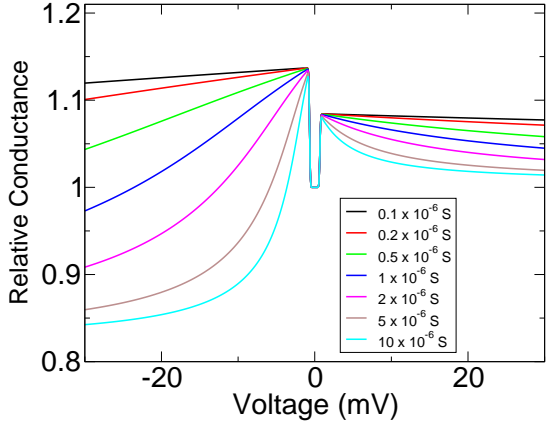}
\caption{
Relative conductance of the Mn/CuN/Cu(100) system as a function the tip
bias. The tip polarization is $\eta = 0.24$ and the B field, equal to
7T, is along the x-axis. The finite population of the excited states is
taken into account. The various curves correspond to various absolute
conductances at zero bias (0.1, 0.2, 0.5, 1., 2., 5. and 10. $10^{-6}$
S). In the figure, the conductance is plotted in relative value, with the
conductance for zero bias set to 1.  } \label{fig_cond_7T} \end{figure}

\begin{figure}
\includegraphics[angle=0,width=0.4\textwidth]{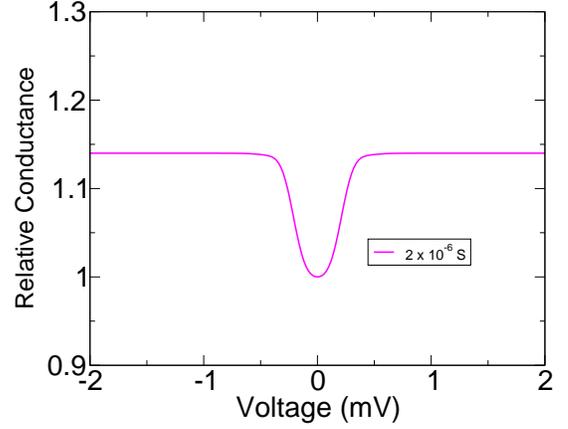}
\caption{
Relative conductance of the Mn/CuN/Cu(100) system as a function the tip
bias. The tip is non-polarized and the B field, equal to 3T, is along
the x-axis. The finite population of the excited states is taken into
account. The conductance is independent of the current. In the figure,
the conductance is plotted in relative value, with the conductance for
zero bias set to 1.  } \label{fig_cond_3T_non_pol} \end{figure}

\begin{figure}
\includegraphics[angle=0,width=0.4\textwidth]{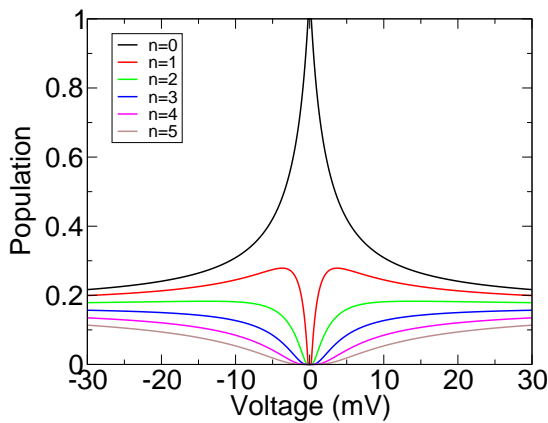}
\caption{
Population of the six magnetic states in the Mn/CuN/Cu(100) 
system as a
function of the tip bias for a non-magnetic tip. 
The magnetic field is along the x-axis and equal
to 3T. The conductance
at zero bias is equal to $2. 10^{-6}$ S.  In contrast
to Fig.~\ref{fig_populations_pol}, there is no bias asymetry. }
 \label{fig_Pop_3T_non_pol} \end{figure}

Figures \ref{fig_cond_3T} and \ref{fig_cond_7T} present the conductivity
of the Mn junction computed with a finite excited state population
for different values of the conductance at $V=0 : 0.1, 0.2 , 0.5,
1., 2.,  5.$ and $10.$ (all in $10^{-6}$ S). They correspond to a B
field of 3 and 7 T, resp. and a tip polarisation of 0.24. The computed
conductivity as a function of V has been convoluted with a Gaussian of
0.2 meV width to mimic the various broadening phenomena. Note that the
actual height of the inelastic conductivity step for B = 3 T depends
on the magnitude of the broadening effect, due to the small excitation
energies (Fig.~\ref{fig_Mag_En_Lev}). The behaviour is the same as
observed experimentally: \textit{i)} there is a sharp peak at very
low voltage corresponding to the excitation thresholds; \textit{ii)}
the conductivity drops as $|V|$ is increased due to the excited state
population; \textit{iii)} this drop is steeper and deeper on the $V <
0$ side;  \textit{iv)} the steepness of the conductivity drop decreases
as the tunneling current decreases to practically vanish for very small
conductances and \textit{v)} the conductance drop is steeper at 3T than
at 7 T due to a longer lifetime of the excited states at 3T. The height
of the inelastic steps and their asymmetry around zero, as well as the
extent of the decrease of the conductance as $|V|$ increases depends on
the polarisation of the tip. In the case of an unpolarised tip ($\eta=0$),
Fig. \ref{fig_cond_3T_non_pol} shows that, in the same way as for the
experimental results, the conductivity is symmetric for $V > 0$ and $V <
0$ and it is practically independent of the tunneling current, although
a significant population of excited states is present. 
Figure \ref{fig_Pop_3T_non_pol} shows the excited state population as
a function of the junction bias for a conductivity of 2.0 $10^{-6}$
S at $V = 0$ and for an unpolarized tip. For large bias, the excited
state populations are large (their sum is larger than the population
of the ground state); they are equal for positive and negative
biases. Though no effect of these excited state populations is apparent
in Fig. \ref{fig_cond_3T_non_pol}. This is a direct consequence
of Fig.~\ref{fig_relat_cond_non_pol} where the conductances of all the
magnetic states are equal above the excitation thresholds in the case of
unpolarised electrons (or holes). Differences could only appear in the
voltage region in between the inelastic thresholds, but 
in the present case, due to the smallness of the  difference between
the various excitation energy threshols, they are hidden by broadening
effects. All these qualitative features nicely agree with the experimental
observations.  The two sets though differ quantitatively, coherently with
the decay rate comparison in Fig. \ref{fig_decay_rates_comp}. The computed
lifetimes are shorter and consequently, the effect of the excited state
population on the junction conductivity appears for larger conductances
and larger tunneling currents.

 The effect of the populations of excited states on the polarized
conductance is much visible in Figs. \ref{fig_cond_3T} and
\ref{fig_cond_7T} which show the conductance as a function of bias for
a fixed tip position. They also strongly emphasize a dependance of the
conductance on the conductuctivity at $V=0$. One can stress that if all
the curves in e.g. Fig. \ref{fig_cond_3T} are plotted as a function
of the junction current $intensity$, then they look very much alike,
the population of the excited states depending dominantly on the current
intensity . In such a plot, the role of the bias voltage and conductivity
at $V = 0$ only appear  in the excitation threshold region.

\subsection{Symmetry of the tunneling process}

\begin{figure}
\includegraphics[angle=0,width=0.4\textwidth]{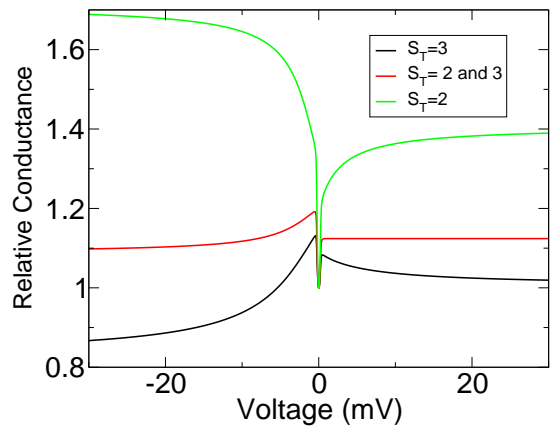}
\caption{
Relative conductance of the Mn/CuN/Cu(100) system as a function of the
tip bias for various hypothesis in the spin-coupling scheme for the
tunneling electrons ($S_T = 3$ is the coupling scheme predicted by
our DFT calculations). The tip polarization is $\eta = 0.24$ and the B
field, equal to 3 T, is along the x-axis. The absolute conductance is
equal to $2.10^{-6}$ S for zero bias. In the figure, the conductance is
plotted in relative value, with the conductance for zero bias set to
1. The finite population of the excited states is taken into account.
} \label{fig_relat_cond_schemes} \end{figure}

A very important feature in our strong coupling treatment of magnetic
transitions is the symmetry of the tunneling state, i.e. which is the
value of $S_T$  that is dominating tunneling through the adsorbate
(see section~\ref{spin_dep_cond} above and a more detailed account
in Ref. [\onlinecite{Gauyacq10}]). As shown by our DFT study in
Ref.[\onlinecite{Lorente09}], the $S_T = 3$ symmetry is dominating in
the case of Mn/CuN/Cu(111). Different values of $S_T$ lead to different
strengths of the magnetic transitions. For illustrative purpose,
we checked the effect of changing the symmetry of the tunneling
intermediate in the present study. Figure ~\ref{fig_relat_cond_schemes}
shows the conductivity as a function of the STM bias, V, for a B field
of 3T, a tip polarisation of $\eta = 0.24$ and a conductance at $V = 0$
of $2. 10^{-6}$ S for three different tunneling symmetry hypothesis:
\textit{i)} $S_T = 3$ dominating, \textit{ii}) $S_T = 2$ dominating and
\textit{iii)}  both $S_T = 3$ and 2 contributing in a statistical way.
This change of symmetry only concerns the transitions induced by the
tunneling electrons and not the excited state decay by electron-hole
pair creation. Not surprisingly, the present physical situation with
a partially polarised tip is highly sensitive to the symmetry of the
tunneling intermediate. The three different symmetries are seen to lead
to different qualitative behaviours: the heights of the inelastic steps at
small  $|V|$ are changing with the symmetry of the tunneling intermediate
indicating a different efficiency of the various $S_T$ symmetries in the
magnetic excitation process. In addition, depending on the symmetry and on
the sign of $V$, the effect of excited state populations is seen to lead to
an increase, a decrease or to a flat behaviour of the conductivity. Only
the results for $S_T = 3$ do exhibit the qualitative behaviour found
experimentally. This comparison strongly supports our assignment of the
$S_T = 3$ symmetry as the dominant symmetry for the tunneling process in
the present system. It also stresses the importance of considering the
proper symmetry and proper spin coupling of the tunneling process for accounting for inelastic
magnetic tunneling, especially with the present experimental protocol
that involves a spin polarised tip and is thus much more selective.

\section{Concluding summary}

We reported on a theoretical study of the lifetime of excited magnetic
states on single Mn adsorbates on a CuN/Cu(100) surface. Electrons
tunneling through a single adsorbate are very efficient in inducing
magnetic transitions as was revealed by IETS experiments; similarly,
substrate electrons colliding on adsorbates are also very efficient in
quenching a magnetic excitation in a single adsorbate on a surface,
the magnetic excitation energy being transferred into a substrate
electron-hole pair. The present study on the quenching process is
inspired by a recent study of magnetic excitations induced by tunneling
electrons~\cite{Lorente09,Gauyacq10}. It is based on a DFT calculation
of the system associated to a strong coupling approach of the spin
transitions. We show that the decay rate of an excited magnetic state
can be written as the product of the flux of substrate electrons hitting
the adsorbate by a spin transition probability. In the Mn/CuN/Cu(100)
system, the lifetime of the excited states is found to be rather short,
the calculated lifetimes are typically in the 0.04-0.4 ns range  for an
applied magnetic field in the 0-7 T range. The lifetime decreases when
the B field rises.

For the comparison with Loth {\em et al.} experiments~\cite{Loth10}, we performed
a detailed study of electron tunneling through the adsorbed Mn atom
in the case of a polarized tip. Using polarized electrons (or holes)
unveils several phenomena among which we can stress: 
\begin{itemize}
 \item As shown in our strong coupling approach, the selection introduced
by the symmetry of the total spin state in the tunneling process (total
spin = spin of the tunneling electron + spin of the adsorbate) is very
important and this leads to large differences in conductance between
various situations: tunneling electrons (or holes) with spin up (or
down), adsorbates in different magnetic states. All these can be related
to the existence of a dominant spin symmetry in the tunneling process,
which is $S_T=3$ in the case of Mn on CuN/Cu(100) as shown here.
 \item As the excitation probability by tunneling electrons is large,
significant populations of excited states can exist for a finite
current. Very interestingly, this leads to observable changes in the
conductance as a function of current in the case of a polarized tip but
not in the case of a non-polarized tip. This is a direct consequence
of one of the key properties of the magnetic excitation process, as
described in the strong coupling approach. The excitation process appears
as a sharing process: a global tunneling current is shared among the
various magnetic states, using sharing probabilities obtained from spin
coupling coefficients. If we sum over all possible spin directions for the
tunneling electron, these spin coupling coefficients sum to one. Thus,
with a non-polarized tip, the conductance for biases above all inelastic
thresholds is independent of the initial state and then independent of
a finite stationary population of excited states. In contrast, in the
case of a polarized tip, the sum over electron spin directions is not
complete and the conductance depends significantly on the magnetic state.
\item The present theoretical results reproduce all the
behaviours observed experimentally, basically those are consequences
of the two phenomena described above. Quantitatively, although the
excitation process is perfectly accounted for, the computed lifetimes of
the excited states appear to be a factor 3 shorter than those extracted
from experiment. As a consequence,  the experimentally observed variation
of the junction conductivity with the tunnelling current is reproduced
but for larger tunnelling currents.  \end{itemize}

\begin{acknowledgments}
F.D.N acknowledges support from Spain's MICINN Juan de la Cierva
program. Computing resources from CESGA are gratefully acknowledged.
Financial support from the spanish MICINN through grant FIS2009-12721-C04-01
is gratefully ackowledged.
\end{acknowledgments}

\end{document}